# Critical analysis of the Bennett–Riedel attack on the secure cryptographic key distributions via the Kirchhoff-law–Johnson-noise scheme

Laszlo B. Kish [1,*], Derek Abbott [2], Claes G. Granqvist [3]

[1] Department of Electrical and Computer Engineering, Texas A&M University, College Station, TX 77843-3128, USA   [2] School of Electrical and Electronic Engineering, University of Adelaide, Adelaide, South Australia, Australia   [3] Department of Engineering Sciences, The Ångström Laboratory, Uppsala University, P. O. Box 534, SE-75121 Uppsala, Sweden

# Critical analysis of the Bennett–Riedel attack on secure cryptographic key distributions via the Kirchhoff-law–Johnson-noise scheme

Laszlo B. Kish [1,*], Derek Abbott [2], Claes G. Granqvist [3]

[1] Department of Electrical and Computer Engineering, Texas A&M University, College Station, TX 77843-3128, USA

[2] School of Electrical and Electronic Engineering, University of Adelaide, Adelaide, South Australia, Australia

[3] Department of Engineering Sciences, The Ångström Laboratory, Uppsala University, P. O. Box 534, SE-75121 Uppsala, Sweden

## Abstract

Recently, Bennett and Riedel (BR) (http://arxiv.org/abs/1303.7435v1) argued that thermodynamics is not essential in the Kirchhoff-law–Johnson-noise (KLJN) classical physical cryptographic exchange method in an effort to disprove the security of the KLJN scheme. They attempted to demonstrate this by introducing a dissipation-free deterministic key exchange method with two batteries and two switches. In the present paper, we first show that BR's scheme is unphysical and that some elements of its assumptions violate basic protocols of secure communication. All our analyses are based on a technically-unlimited Eve with infinitely accurate and fast measurements limited only by the laws of physics and statistics. For non-ideal situations and at active (invasive) attacks, the uncertainly principle between measurement duration and statistical errors makes it impossible for Eve to extract the key regardless of the accuracy or speed of her measurements. To show that thermodynamics and noise are essential for the security, we crack the BR system with 100% success via passive attacks, in ten different ways, and demonstrate that the same cracking methods do not function for the KLJN scheme that employs Johnson noise to provide security underpinned by the Second Law of Thermodynamics. We also present a critical analysis of some other claims by BR; for example, we prove that their equations for describing zero security do not apply to the KLJN scheme. Finally we give mathematical security proofs for each BR-attack against the KLJN scheme and conclude that the information theoretic (unconditional) security of the KLJN method has not been successfully challenged.

## Introduction

Information theoretic (*i.e*., unconditional) security [1] means that the stated security level—either perfect or imperfect, as in any physical system [2]—holds even for cases when the abilities of an eavesdropper (generally called "Eve") are limited only by the laws of physics. Since 1984, quantum key distribution (QKD) [2] has been claimed to possess unconditional security and much later, in 2005, an alternative based on classical physics, known as the Kirchhoff-law–Johnson-noise (KLJN) scheme [2], appeared as a competing approach.

Very recently, QKD's co-founder Charles Bennett [3] co-authored a manuscript [4] with Jess Riedel wherein they present an extensive criticism of the KLJN scheme and deny its security under idealized conditions. Bennett and Riedel (BR) assert that thermodynamics is not essential in the KLJN scheme and argue that it does not provide security. They attempt to prove this claim by showing a dissipation-free deterministic key exchange method with nothing but two batteries and two switches. Moreover, among other statements [4], BR argue that the quasi-stationary (*i.e.*, no-wave) limit of electrodynamics is unsuited for information transfer, thus implying that this (required) assumption [2] for (perfect) security of the KLJN system is unphysical. Our present paper is a detailed critical analysis of the BR scheme. In summary, we show that BR's scheme is unphysical, and we provide further analysis that demonstrates the security of the KJNL scheme.

In this introductory chapter we set the scene for the next chapter, wherein we will fully crack the BR system in various ways and also respond to BR's arguments about the KLJN scheme. We first consider the currently ongoing debates concerning the security of QKD, which is a necessary preamble since BR propound that the security of QKD is robust. Then we briefly outline the KLJN secure key distribution scheme and its main features. Subsequently, we describe the "thermodynamics-free" key exchange system due to BR and the related argumentation in their paper [4].

**1.1 Is the security of quantum encryption indeed robust?**

BR write [4]: "*we emphasize that quantum key distribution has been shown to be robust with imperfect components against very general attacks*". We see this situation very differently and first briefly summarize the currently ongoing debates in the QKD field.

Currently, there is a discussion [5–8] about the fundamental security/non-security of existing QKD schemes. This debate was initiated by Yuen [5,8], who was later joined by Hirota [6] in claiming that the security of existing quantum key distribution schemes is questionable or poor. Recently, Renner [7] entered the discussion to defend the old security claims. It should be noted that Yuen [9] and Zubairy *et al.* [10] have proposed new advanced schemes for non-QKD-based secure quantum communication.

BR's claim that QKD displays "*robust security with imperfect elements*" [4] has been proven incorrect, and QKD has been cracked by utilizing the imperfect nature, such as non-linearity, of necessary building elements. Practical quantum communicators—including several commercial ones—have been fully cracked as shown in numerous recent papers [11–25]. Vadim Makarov, who is one of the leading quantum crypto crackers, stated that "*Our hack gave 100% knowledge of the key, with zero disturbance to the system*" [11]. This statement hits the foundations of quantum encryption schemes, because the often-claimed basis of the security of QKD protocols is the assumption that any eavesdropping activity will disturb the system enough to be detected by the communicating parties (generally referred to as "Alice" and "Bob"). An important aspect of these quantum-based hacking attacks is the extraordinary (100%) success ratio of extracting the "secure" key bits by Eve, which indicates that the security is not only imperfect but simply non-existing against these types of attacks until proper defense strategies or protocol modifications have been added to the scheme in order to restore the information theoretic security they supposedly had before these attacks were known.

In conclusion, and in clear contradiction to BR's claim [4], *quantum key distribution has been found vulnerable to well-designed attacks* for the case of imperfect components.

## 1.2 The KLJN secure key exchange system

The Kirchhoff-law–Johnson-noise key distribution scheme [2,26–39] is a classical statistical physical alternative to QKD, whose security is based on Kirchhoff's Loop Law and the Fluctuation-Dissipation Theorem. More generally, it is founded on the Second Law of Thermodynamics, which certifies that the security of the ideal KLJN scheme is as strong as the impossibility to build a perpetual-motion machine of the second kind. Potential and unique technical applications of the KLJN scheme include non-counterfeitable hardware keys and credit cards via Physical Uncloneable Functions (PUFs) [35]; unconditionally secure hardware, computers and other instruments [35,36]; and unconditionally secure smart grids [37–39]. The short summary of the KLJN scheme given below is based on a previous survey paper [2].

### 1.1.1 The idealized KLJN scheme and its security

The working principle of the KLJN scheme [2,26] is presented in Fig. 1, which shows an idealized configuration without any defense circuitry—such as current-voltage measurement/comparison, filters, *etc*—against invasive and non-ideality attacks. At the beginning of each bit exchange period (BEP), Alice and Bob connect their randomly chosen resistors $R_A$ and $R_B$, respectively, to the wire line. These resistors are randomly selected by the switches from the set $\{R_L, R_H\}$, $(R_L \neq R_H)$, where the elements represent the low *L* and high *H* bit values 0 and 1, respectively.

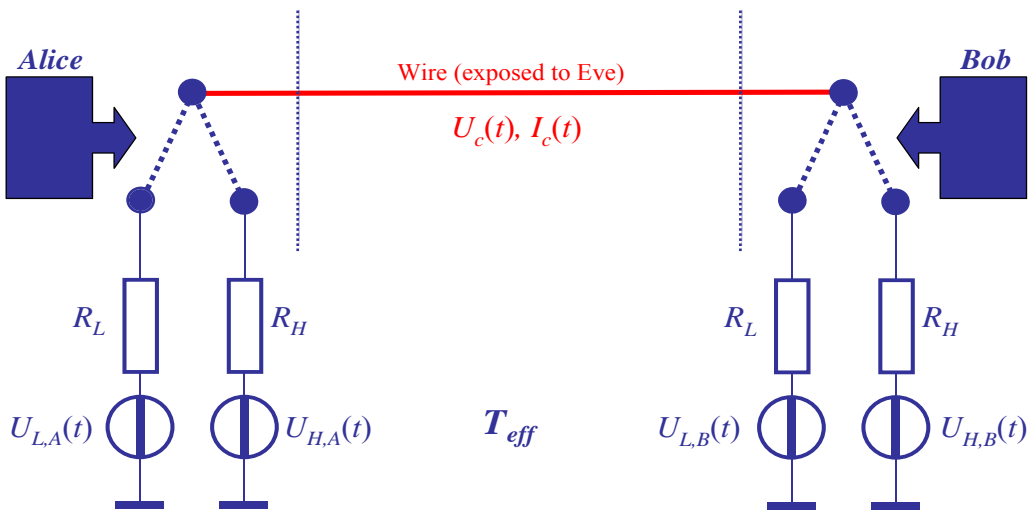

**Figure 1. Outline of the core KLJN key exchange system.** The communicator parties, Alice and Bob, randomly choose and connect either $R_L$ or $R_H$ to the wire. The (effective) temperature $T_{eff}$ is publicly agreed and kept, and the (enhanced or standard) Johnson noises of the resistors $U_{L,A}(t)$, $U_{L,B}(t)$, $U_{H,A}(t)$, and $U_{H,B}(t)$ are independent and Gaussian. The resulting channel voltage $U_c(t)$ and current $I_c(t)$ are also uncorrelated due to the Second Law of Thermodynamics. Parasitic elements leading to non-ideal features and defense circuitry against active (invasive) attacks and against attacks utilizing non-ideal features are not shown.

The Gaussian voltage noise generators—delivering white noise with publicly agreed bandwidth—represent an enhanced thermal (Johnson) noise at a publicly agreed high effective noise-temperature $T_{eff}$ at which their noises are statistically independent from each other, implying that $\langle U_A(t)U_B(t)\rangle = 0$, as well as from the noise during a former BEP. During the first practical implementation of the KLJN scheme, by Mingesz *et al.* [29], the noise-temperature range $8 \times 10^8 \text{K} \leq T_{eff} \leq 8 \times 10^{11} \text{K}$ was used, which made the wire temperature insignificant even when the wire resistance was not zero.

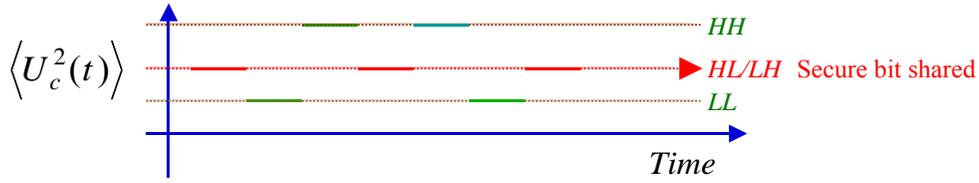

**Figure 2. Mean-square voltage (and current) versus time during operation.** There are three different levels (dotted lines) depending on the actual bit values; the intermediate value indicates a secure bit exchange period.

Alice and Bob (as well as Eve) can use a measurement of the mean-square voltage and/or current to assess the bit status of the system, as shown in Fig. 2 for the case of voltage. The situations *LH* and *HL* represent secure bit exchange [2,26], because Eve cannot distinguish between them through measurements, and whenever Alice and Bob see the HH/LH situation they know that the other party has the complementary bit value, which means that they infer the full bit arrangement. Eve cannot extract this information, because she does not know any of the bit values. In other words, a secure bit has been generated and shared. The bit situations *LL* and *HH* are insecure, which means that these bits (50% of the executed BEPs) are discarded by Alice and Bob.

According to the Fluctuation-Dissipation Theorem, the power density spectra $S_{u,L}(f)$ and $S_{u,H}(f)$ of the voltages $U_{L,A}(t)$ and $U_{L,B}(t)$, supplied by the voltage generators in $R_L$ and $R_H$, are given by

$$S_{u,L}(f) = 4kT_{eff}R_L \quad \text{and} \quad S_{u,H}(f) = 4kT_{eff}R_H, \tag{1}$$

respectively. In the case of secure bit exchange (*i.e.*, the *LH* or *HL* situation), the power density spectrum $S(f)$ and the mean-square amplitude $\langle U_{ch}^2 \rangle$ of the channel voltage $U_{ch}(t)$, and the same measures of the channel current $I_{ch}(t)$, are given by

$$\langle U_{c,HL/LH}^2 \rangle = \Delta f \, S_{u,c,HL/LH}(f) = 4kT_{eff} \frac{R_L R_H}{R_L + R_H} \Delta f, \tag{2}$$

and

$$\langle I_{c,HL/LH}^2 \rangle = \Delta f \, S_{i,c,HL/LH}(t) = \frac{4kT_{eff}}{R_L + R_H} \Delta f, \tag{3}$$

respectively, where $\Delta f$ is the noise bandwidth.

**1.1.2 The security of the KLJN scheme is based on the Second Law of Thermodynamics**

During the *LH* and *HL* cases, linear superposition makes the spectrum given by Eq. (2) represent the sum of the spectra at the two particular situations. Thus one obtains

$$S_{L,u,c}(f) = 4kT_{eff} R_L \left( \frac{R_H}{R_L + R_H} \right)^2 \tag{4}$$

when only the noise generator due to $R_L$ is running and

$$S_{H,u,c}(f) = 4kT_{eff} R_H \left( \frac{R_L}{R_L + R_H} \right)^2 \tag{5}$$

when the only the noise generator due to $R_H$ is running.

If Eve is to identify which end of the wire has $R_L$ or $R_H$, it is necessary for her to measure and evaluate a physical quantity offering directional information. In the ideal case, the only information of this kind is the direction of the power flow from Alice to Bob (or *vice versa*, depending on the choice of positive current direction). In thermal equilibrium, however, this power must fulfill $P_{A \to B} = \langle U_c(t) \, I_c(t) \rangle = 0$, as required by the Second Law of Thermodynamics. In other words, the ultimate security of the KLJN system against passive attacks is provided by the fact that the power $P_{H \to L}$, by which the noise generator due to resistor $R_H$ is heating resistor $R_L$, is equal to the power $P_{L \to H}$ by which the noise generator due to resistor $R_L$ is heating resistor $R_H$ [2,26,32]. Thus the fact that the net power flow is governed by $P_{A \to B} = P_{L \to H} - P_{H \to L} = 0$ can easily be shown from Eqs. (4) and (5) for the noise-bandwidth $\Delta f$

by

$$P_{L \to H} = \frac{S_{L,u,c}(f)}{R_H} = 4kT_{eff} \frac{R_L R_H}{(R_L + R_H)^2} \Delta f \tag{6a}$$

and

$$P_{H \to L} = \frac{S_{H,u,c}(f)}{R_L} = 4kT_{eff} \frac{R_L R_H}{(R_L + R_H)^2} \Delta f \tag{6b}$$

The equality $P_{H \to L} = P_{L \to H}$ is in accordance with the Second Law of Thermodynamics. In other words it is as difficult to crack the ideal KLJN scheme as to build a perpetual motion machine of the second kind [4].

This security proof against passive (listening) attacks holds only for Gaussian noise—*i.e.*, the statistics of thermal noise—which has the well-known property that its power density spectrum or autocorrelation function already provides the maximum achievable information about the noise, and no higher-order distribution functions or other tools, such as higher-order statistics, are able to provide additional information.

The required duration $\tau$ of the BEP, at a given bit error probability [34] of the bit exchange between Alice and Bob, is determined by the following arguments: For the *LL* bit status of Alice and Bob, which is not a secure situation, the channel voltage and current satisfy

$$\langle U_{c,LL}^2 \rangle = \Delta f\, S_{u,c,LL}(f) = 4kT_{eff} \frac{R_L}{2} \Delta f \quad \text{and} \quad \langle I_{c,LL}^2 \rangle = \Delta f\, S_{i,c,LL}(t) = \frac{2kT_{eff}}{R_L} \Delta f, \tag{7}$$

while, in the case of the other non-secure situation namely the *HH* bit status, the channel voltage and current satisfy

$$\langle U_{c,HH}^2 \rangle = \Delta f\, S_{u,c,HH}(f) = 4kT_{eff} \frac{R_H}{2} \Delta f \quad \text{and} \quad \langle I_{c,HH}^2 \rangle = \Delta f\, S_{i,c,HH}(t) = \frac{2kT_{eff}}{R_H} \Delta f. \tag{8}$$

During key exchange in this classical way, Alice and Bob must compare the predictions of Eqs. (7) and (8) with the actually measured mean-square channel voltage and current to decide whether the situation is secure (*i.e.*, *LH* or *HL* prevails), while realizing that these mean-square values are different in each of these three situations (*LL*, *LH* or *HL*, and *HH*). If the situation is secure, Alice and Bob will know that the other party has the inverse of his/her bit, which implies that a secure key exchange takes place. Alice and Bob must use sufficiently large statistics to achieve low error probability. Fortunately, the bit error probability decays exponentially with the duration $\tau$ of the BEP [34]. Furthermore, a new "intelligent" KLJN protocol [31] can be used, which employs additional circuit calculations by Alice and Bob to reduce the BEP without increasing the error probability.

**1.1.3 On active (invasive) attacks and attacks utilizing non-idealities**

It has been pointed out repeatedly [2,26,28,29,32] that deviations from the earlier shown circuitry and Johnson-like noise—including invasive attacks by Eve, parasitic elements, delay effects, inaccuracies, non-Gaussianity of the noise, *etc*—will cause a potential information leak toward Eve. However it is fortunate that the KLJN system is very simple, which implies that the number of such attacks is strongly limited. The defense methods against the attacks are straightforward and are generally based on the comparison of instantaneous voltage and current data at the two wire ends via an authenticated communication between Alice and Bob, as indicated in Fig. 3. These attacks [2,40,43,45] are not the subject of the present paper, and we refer to our relevant rebuttals where they have been analyzed [2,32,41,42,44] and where misconceptions and errors have been pointed out and corrected. Our earlier survey paper [2] reviewed various attacks on the KLJN scheme.

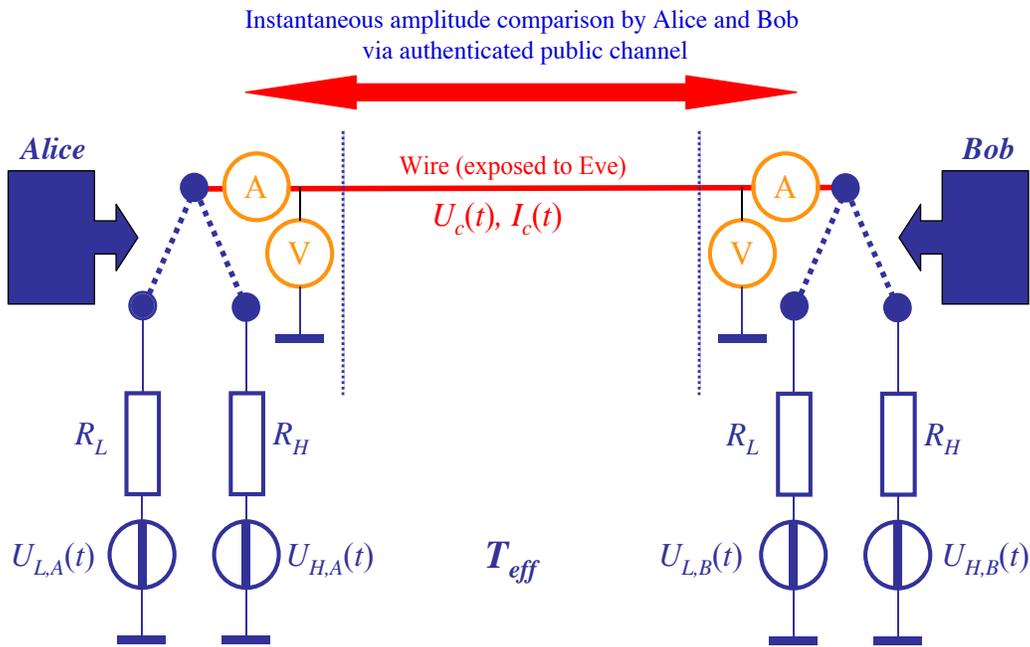

**Figure 3. KLJN system minimally armed against invasive (active) attacks, including the man-in-the-middle-attack.** Alice and Bob measure the instantaneous channel voltage and current amplitudes and compare them via an authenticated public channel. In this way, they learn all the information Eve can have. Additions to prevent hacking—such as line filters, blinding detectors, *etc*—are not shown. The notation is the same as in Fig. 1.

It is important to emphasize that Alice and Bob know Eve's best measurement information, because it is given by comparisons of voltage and current at the two ends of the wire. If Eve uses the best available protocol and the security of a certain bit is compromised, this is known also by Alice and Bob who therefore can decide to discard the bit in order to have a secure key. This is a new and unique situation in cryptography, which raises a number of research questions as mentioned in an earlier paper [32].

Finally, a secure type of privacy amplification [33]—XOR-ing the key bit pairs and producing a new key with this output, which results in half of the original length—is also feasible to enhance the security because of the low bit error probability of KLJN key exchange. The error probability decays exponentially with the increasing duration $\tau$ of the BEP [34]. At the experimental demonstration [29] the error probability was $2 \times 10^{-4}$.

**1.1.4 Foundations of the information-theoretic security in practical KLJN schemes**

Of course, perfect security of any physical key exchanger exists only under ideal (mathematical) conditions. For example, quantum encryption theoretically can offer perfect security only in the limit of a zero-photon-emission rate [5] (*i.e.*, zero bit exchange rate) and zero detector and channel noise limits, which are unphysical and can never occur in a real system. The KLJN scheme is no exception to this rule [2,26,32]: it offers perfect security only at zero bandwidth or distance as a consequence of transients, cable resistance, capacitance, *etc*. However, just as for claims in favor of QKD, parameters of the KLJN building elements and protocol can be chosen so that the perfect security limit can be approached asymptotically. The general situation in the non-ideal case is that a miniscule DC signal component buried in a much larger Gaussian noise (of fixed variance) must be detected by Eve from small statistics limited by the BEP. This DC signal component is typically the mean value of a finite-time mean-square operation or that of the output component of a cross-correlation operation; further discussion on this issue is given in an earlier paper [31] and in Sec. 2.8. Eve must detect the sign of this small DC component in the large noise. When the parameters approach the ideal situation, the ratio of the DC signal amplitude and the root-mean-square (RMS) amplitude of the noise converges towards zero as a power law decay—typically with exponent –1 or –2 [2,29,42,44]—with regard to the invested resources such as wire volume, current/voltage resolution, BEP duration $\tau$, *etc*.

Here we summarize the foundation of the mathematical analyses below by explaining why and how Eve's information is limited compared to Alice and Bob's information, even though we have not imposed any limitation for Eve's measurement accuracy or her measurement speed anywhere in our analysis. This is a point of frequent misunderstanding and the additional explanation here was triggered by the referee report written by Dr. Bennett.

All our analyses and argumentations are based on a *technically-unlimited eavesdropper* (Eve). Accordingly, Eve's amplifiers and analyzers have infinite amplitude resolution and infinite measurement speed (infinite bandwidth), they are noise-free and absolutely linear. Otherwise, parameters describing characteristic time/frequency cut-off of Eve's measurement apparatus and of the amplitude resolution limit would enter into all of the equations describing Eve's information. Eve's only measurement-technology limitation is of fundamental physical nature: Eve can only very poorly separate the signals originating from the two directions as the result of the Rayleigh scattering at sizes much smaller than the signal wavelength; see Sections 2.1.4 and 2.3.2 below. This fundamental physical process is unavoidable, and Eve must live with the fact that the laws of physics prohibit using an efficient directional coupler. However, the same limitations concerning Rayleigh scattering also apply to Alice and Bob.

Then, the obvious question emerges: *Why does Eve have significantly less information than Alice and Bob*? One should note that this question is relevant only for passive attacks in practice (that is, non-ideal situations with finite distance, finite bandwidth, nonzero cable resistance and capacitance) and active attacks for arbitrary situations, because, for the ideal situation (with zero distance, finite bandwidth, zero cable resistance and capacitance) the Second Law guarantees perfect security. So, how can we summarize the main arguments behind our security proofs against the various attacks presented in this paper, and what is the fundamental mechanism behind the unconditional security shown by our mathematical security proofs?

Is it today's technological limitations of Eve's accuracy? No: it is a *fundamental limit*: the *statistical uncertainty principle* interrelating the finite measurement duration with statistical errors during noise analysis.

Eve can measure with infinite speed, which means infinite bandwidth, but *she will not find anything beyond the noise bandwidth* $\Delta f$ set by the noise generators and line filters of Alice and Bob (except negligible spurious, stochastic frequency components decaying in an exponential or power-law fashion versus frequency beyond the band limit). According to Shannon's sampling theorem, the maximum sampling frequency giving statistically independent data is $2\Delta f$ thus during the duration $\tau$ of the BEP, Eve (just as Alice and Bob) can extract only $s = 2\Delta f \tau$ independent samples of the channel noise even if Eve's measurement apparatus can collect samples with infinite sampling frequency. The quantity $s$ is determined by Alice and Bob because $\Delta f$ and $\tau$ are determined by them and chosen so that the error probability of exchanged bits is miniscule, thus no error correction algorithm is needed. Yuen [5] points out that error correction algorithms provides information to Alice and Bob in QKD, but KLJN can avoid using such. In practical applications one has $s \approx 100$, which is an extraordinarily small number for Eve to safely distinguish the minor differences, $10^{-4}$ or less [41], in the noise statistics she can extract using non-idealities. For active attacks, Alice and Bob can easily enforce the same small differences in the statistics—for example 14 bits, used in the experimental demonstration, which is equivalent to a difference of less than $10^{-4}$. It is virtually impossible to distinguish such a small difference between stochastic signals with the available number $s$ of independent samples. As a result, the mathematical analyses shown in the subsequent sections indicate that Eve's successful guessing probability will be close to 0.5,—i.e., the limit representing zero information—and the accepted measure of security (statistical distance, see below) between Eve's extracted version of the key and a perfectly secure key is exponentially small.

*But what about Alice and Bob*? Alice and Bob know essential parameters that Eve does not have access to: they know their own resistance value and the exact amplitude of their noise fed into the line. Thus they do not need to utilize the non-ideality-based miniscule differences seen by Eve or the even smaller differences that Eve may generate by active (invasive) attack while staying hidden; they only need to monitor the channel voltage/current and identify which one of the three significantly different levels of the mean-square noise takes place. At $s \approx 100$, the achievable bit exchange has extremely small error probability, such as $10^{-12}$, see [34]. Further improvements are offered by the "intelligent" KLJN method, where $s$ can be significantly diminished by Alice and Bob via reducing $\tau$ when utilizing the knowledge of their own noise time function, combined with linear network calculations [31] allowed by the classical physical nature of the scheme.

**1.1.5 Mathematical proof of the unconditional security of the exchanged key**

In order to mathematically analyze the security of the shared key, one must compare the probability distribution for successfully guessing each possible key sequence of an *N*-bit-long key, encompassing $2^N$ different sequences, with that of the perfect key having uniform distribution. A statistical distance measure, the variational distance $\Delta$ [46] between the distributions representing the key guessed by Eve and the distribution representing the ideal (uniform) key is a useful concept. It defined by

$$\Delta(E,I) = \max_{j=1,\ldots,2^N} \left[ P(E_j) - P(I_j) \right] , \qquad (9)$$

where *E* and *I* represent Eve's extracted key and the perfect key, respectively, and $P(E_j)$ and $P(I_j)$ are the probabilities of correctly guessing the $j^{\text{th}}$ version of Eve's key and of the perfect key, respectively. The key exchange has *ε*-security, as discussed by Hirota [6], if the statistical distance between the distributions representing the key guessed by Eve and the ideal (uniform) keys is less than $\varepsilon$, *i.e.*,

$$\Delta(E,I) \leq \varepsilon \qquad (10)$$

for $\varepsilon \geq 0$.

The KLJN scheme provides identically and independently distributed sequences of random variables as key bit values, so that

$$\Delta(E,I) = \max_{j=1,\ldots,2^N} \left[ P(E_j) - P(I_j) \right] = p^N - 0.5^N , \qquad (11)$$

where *p* is Eve's probability of successfully guessing bits. In non-ideal cases involving an information leak, and when the parameters are sufficiently close to the ideal limit, *p* can be given as

$$p = 0.5 + q , \qquad (12)$$

where $0 < q \ll 0.5$; here $q = 0$ would mean a perfectly secure key. The reason for this behavior is easy to see if one realizes that Eve's small DC signal component offsets the center (mean value) of the probability density burying the large Gaussian noise. The first derivative at the center of the Gaussian density function is zero, implying that its Taylor approximation, to first order, results in a stable value for small changes around the center. In the idealized case (*i.e.*, zero DC signal) Eve's estimation of the mean value of the Gaussian noise would yield $p = 0.5$ (recording a positive sign at 50% of the exchanged key bits and negative sign also at 50% of the cases). In the non-ideal situation, the DC signal and the mean value of noise + signal are positive or negative; hence the flat amplitude distribution within this range makes Eve experience a non-zero $q$ (*cf.*, Eq. 12) which is proportional to the DC signal [47].

As an example, we now consider the case of a non-zero wire resistance [29,40-42] and assume that capacitive effects are compensated [29] or can be neglected due to the actual bandwidth. More examples will be shown in Chapter 2. For the case of fixed distance and bandwidth, $q$ is proportional to the inverse of the square of wire diameter, *i.e.*, with the inverse of the wire's volume *V*. In other words

$$q = \vartheta_w V^{-1} , \tag{13}$$

where $\vartheta_w$ is a constant valid for a wire-resistance attack. Then, for the case of $Nq \ll 0.5$, one obtains

$$\Delta = (0.5 + q)^N - 0.5^N = 0.5^N \left[ (1 + 2q)^N - 1 \right] \cong 2Nq 0.5^N = 2N\vartheta_w V^{-1} 0.5^N , \tag{14}$$

where the last approximation is valid for $q \to 0$. Equation (14) indicates that $\Delta$ decays exponentially with increasing value of *N* and inversely with wire volume *V*.

For the case of $\varepsilon$-security with $\Delta \leq \varepsilon$, *i.e.*, in the $Nq \ll 0.5$ limit, the required $q$ is given by

$$q(\varepsilon, N) = \frac{\vartheta}{V(\varepsilon, N)} \leq \frac{\varepsilon}{2N} 2^N . \tag{15}$$

During the experimental demonstrations of the KJNL scheme, referred to above [29], it was found that Eve's $q$ equaled 0.025 for secure bit exchange during wire-resistance attacks with a wire resistance being 2% of the loop resistance. For a practical evaluation, let us suppose, that due to additional leaks (transients, cable capacitance, etc.) the actual $q$ is double of that, $q = 0.05$. Due to the $0 < q \ll 0.5$ assumption, in order to use the theory described above, we apply the a privacy amplification described in [33], which keeps the independently and identically distributed nature of the key by XOR-ing pairs of bits in the original key to have a new key with enhanced security and half of the length. Then, repeating this process a second time to obtain the final key with 25% of the original length, the resulting effective $q$ of Eve is $q = 5 \times 10^{-5}$ [33] which allows key lengths *N* up to the order of $10^4$ bits. Equation (11) can then be evaluated with this effective $q$ value. For a 1000-bit-long shared key, it results in $\Delta(E,I)_{1000} = 9.3 \times 10^{-303}$ (*i.e.*, an $\varepsilon$-security with $\varepsilon_{1000} \cong 10^{-302}$); for a 500-bit-long shared key it results in $\Delta(E,I)_{500} = 1.5 \times 10^{-152}$ (*i.e.*, an $\varepsilon$-security with $\varepsilon_{500} \cong 2 \times 10^{-152}$).

Finally, we observe that there are advanced protocols that can enhance the security or limit the required resources in efficient ways while the scaling of $q$ versus the utilized resource (wire volume, *cf*. Eq. 13) shown above in Eqs. (14) and (15) does not change. Below we give a short list of advanced protocols and associated basic security features proposed up to now:

(*a*) Ideal KLJN schemes with passive attacks:

- The Second Law of Thermodynamics and Kirchhoff's Loop Law [2,26].

(*b*) Non-ideal KLJN schemes with passive or active (invasive) attacks:

- Transient protocols involving random-walk from equal resistances [31] and voltage ramping/timing [2,29].
- Selecting the noise bandwidth versus the value of wire resistance and wire capacitance [2,29].
- General defenses that work in any situation including hacking, encompassing comparison of instantaneous voltage and current amplitudes and discarding any bits where they differ or where they provide information to Eve (even it is erroneous). Note, however, that specific protocols apply for different hacking attacks.
- Privacy amplification (XOR-ing key bit pairs [33]).
- Enhanced KLJN protocols, for example the "intelligent" (iKLJN) and "keyed" (KKLJN) methods [31].

**1.1.6 Optional security addition: cap imposed on *q* by Alice and Bob**

We briefly mention an additional security tool provided by the classical physical nature of the KLJN system; see its detailed description and security analysis elsewhere [48]. The fact that Alice and Bob have a public authenticated communication channel for comparing their instantaneous current and voltage data (together with the channel parameters known by them) allows them to access the measurement information of Eve. Thus they can discard those exchanged bits that give out too much information to Eve. In another word, Alice and Bob can impose a strict upper limit on *q* in Eqs. (12) and (15). This additional security tool is useful when the available resources of Alice and Bob are insufficient to diminish the *q* related to the exchanged raw bits and, for some other reasons, they want to avoid privacy amplification to reduce it. This ability of Alice and Bob is another indication that they are in full control of the maximum of statistical information Eve is able to access.

**1.2 Summary of Bennett–Riedel's arguments regarding the KLJN scheme**

BR have presented an extensive analysis [4] that is fundamentally flawed but nevertheless very useful for the purpose of elucidating differences between simplistic or irrelevant model approaches and the physics upon which the KLJN scheme is founded.

An outline of BR's claims reads as follows: It is first stated that the no-wave limit (*i.e.*, quasi-static electrodynamics) is unphysical for signal propagation. Based on this statement, they assert that Eve can separate and measure the "orthogonal" wave components propagating from Alice to Bob and *vice versa*. They also state that the KLJN scheme is deterministic, which means that Eve has a full description of the whole system, including Alice's and Bob's history, if Eve's measurements of the two wave components are limited by nothing but the laws of physics. To support these claims, BR expound that thermodynamics and noise are not essential in the KLJN scheme and that thermodynamics would eradicate determinism as a consequence of fluctuations. Further corroboration of their view is obtained from the construction of a deterministic and thermodynamics-free key exchanger, which looks similar to a KLJN scheme without resistor, and where two of the four noise voltage generators are removed and the remaining ones replaced by batteries with known and identical voltage. Moreover, BR propose a passive correlation-measurement-based attack and an active current-extraction attack against the KLJN scheme.

After briefly describing BR's claims, we critically analyze and refute all of them in Chapter 2, and we also present the physics appropriate for the KLJN scheme.

**1.2.1 Bennett–Riedel's claim concerning no information transfer in a wire in the no-wave (quasi-static) limit**

BR write [4]: "*We believe this no-wave limit is inappropriate and nonphysical for analyzing communication protocols (even as a mathematical idealization) because if propagating waves are excluded there is no way for information to get from Alice's side of the circuit to influence Bob's side, or vice versa.*"

Based on this argument, they assert that Eve can separate and measure the "orthogonal" wave components that propagate from Alice to Bob and *vice versa*.

After surveying the relevant physics facts about waves, directional couplers for signal separation and the no-wave limit in Sec. 2.1, we refute the above argument in Sec. 2.2. Furthermore, we show what physics has to say about signal propagation in the no-wave (quasi-static) limit.

**1.2.2 Bennett–Riedel's claim that the KLJN system does not offer security**

BR [4] set up three equations for the KLJN scheme, which invoke the deterministic nature of Maxwell's equations and neglect the stochastic nature of Johnson noise and the secret/random choice of the resistors. With this premise, it is not surprising that they concluded that KLJN does not offer any security. Here we discuss only the first and third of BR's equations—since the second one is redundant—and their main conclusion.

The conditional information $H(F|G)$ represents the remaining uncertainty about the set of data $F$ when the set of data $G$ is known. Now $H(F|G) = 0$ if $G$ completely determines $F$, whereas

$H(F|G) = H(F)$ for the case when $G$ does not provide any information about $F$. BR's first equation is

$$H(X|Z_A) = H(X|Z_A, Z_B) = H(X|Z, Y),  \qquad (16)$$

where $X$ is a variable that fully describes the physical quantities on Alice's side of Eve's location during the BEP. These quantities include waves traveling toward Alice and away from her and all of her equipment, as well as noise and memory. The variable $Y$ has the same meaning with regard to Bob. Furthermore, $Z_A$ and $Z_B$ are wave components propagating from Alice and Bob (as observed by Eve), respectively, and $Z = (Z_A, Z_B)$ represents both wave components. We note that in BR's paper [4] either $Z$ is incorrectly indexed or $X$ and $Y$ must be exchanged.

We first presume that Eq. (16) is valid, which assumes that $Z_A$ and $Z_B$ can be measured separately. This means that the uncertainty about Alice's "full description" $X$ does not change if Eve expands her knowledge of wave $Z_A$ coming from Alice by the knowledge of wave $Z_B$ coming from Bob, and the same remains true even if knowledge of the total description of Bob's data $Y$ is included.

It should be observed that the first equality in Eq. (16) contradicts BR's proposed passive correlation attack [4], which requires knowledge of both $Z_A$ and $Z_B$ and thus implies that $H(X|Z_A) > H(X|Z_A, Z_B)$.

We now introduce the mutual information $I(X;Y)$ of $X$ and $Y$, which measures how much the knowledge of $X$ or $Y$ tells about the other variable. As a consequence of Eq. (16), and with further argumentation, BR deduce the following equation for the conditional mutual information between $X$ and $Y$, conditional on $Z$:

$$I(X;Y|Z) = H(X|Z) - H(X|Z,Y) = 0. \qquad (17)$$

This equation, if it is valid, would mean that after measuring the two waves $Z = (Z_A, Z_B)$, Eve's information about $X$ (*i.e.*, Alice's full description) is not increased by learning $Y$ (Bob's full description). Thus after measuring the two waves $Z = (Z_A, Z_B)$, Bob's information about Alice would not be larger than Eve's information about her. The same argumentation would work also in the opposite direction, so that the KLJN system would not offer any security.

We will see below that BR's equations are invalid even in the wave limit; this is a result of multiple reflections as well as of Alice's and Bob's secure reflection coefficients and noises (known only by them) that always guarantee that they know more than Eve.

Most importantly, Eqs. (16) and (17) are entirely unfounded in the no-wave limit because the propagating relaxations $Z_A$ and $Z_B$ (which are not waves) cannot be measured separately; only their sum can be determined.

**1.2.3 Bennett–Riedel's claim regarding a "thermodynamics-free" key exchange scheme**

One of the major claims of BR [4] is that thermodynamics and noise are not essential for security in the KLJN scheme. To prove this, they attempted to construct a deterministic key exchange method with two voltage generators and two switches, as illustrated in Fig. 4. This scheme is in fact already known; it is called the "Orlando system" and was conceived and patented by Davide Antilli in 2005 [49]. Despite its origin, we refer to it as the "BR system" below.

In the idle mode between bit exchange periods, the switches are in position *I*; thus the wire channel is grounded. At the beginning of the BEP, Alice and Bob randomly choose between the switch positions *L* or *H* representing the corresponding bit values, and at the middle of the BEP they change their bit value. If the randomly chosen sequences of bit values happen to be identical, then the voltage on the wire will be zero for half of the BEP, and these events are disregarded. If the choices by Alice and Bob are complementary, then the voltage is $U_0$ for the whole BEP.

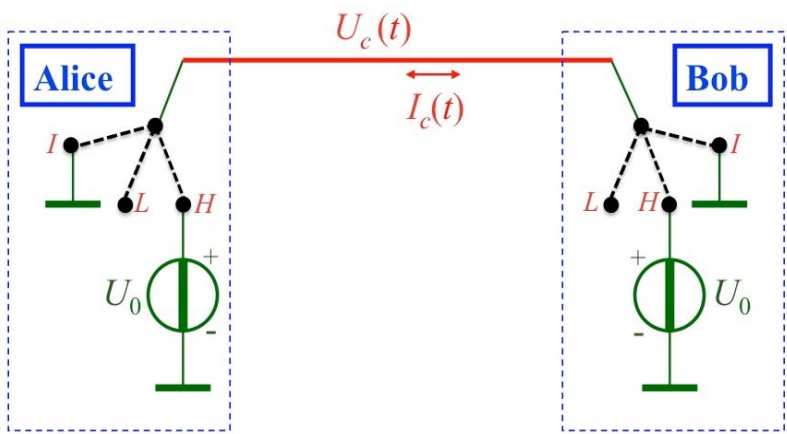

**Figure 4. Outline of the (Antilli–)Bennett–Riedel system.** The two ends of the wire channel are connected to a three-stage switch with positions *I* (idle between two bit exchanges), *L* (low bit value) and *H* (high bit value). The two DC voltage generators have the voltage $U_0$, and $U_c(t)$ and $I_c(t)$ are the voltage and current time-functions in the wire. Note that this figure is an improved version of BR's system because they mention the idle situation (necessarily grounded wire) only in the text but do not show it in their figure.

BR make three statements about the system in Fig. 4, which will be important later: They assert that (*i*) "*The wires and voltage sources are taken to be ideal, with zero thermal noise*" and, as a corollary, that (*ii*) "*Thermodynamics and noise do not play a role.*" Furthermore, they claim (*iii*) that *the BR system is secure in the "no-wave" limit* accomplished in a special way: that Eve waits with her measurements until transients have decayed.

We will see below that statements (*i*) to (*iii*) lead to an unphysical situation, namely that Eve must wait for infinite time before she may start listening. Furthermore, one should note that (*iii*) is an illegal assumption in unconditionally secure communications, because then Eve can only be limited by the laws of physics. Thus statement (*iii*), in itself, would imply only conditional security.

In Secs. 2.1 and 2.2 below we show why BR's scheme is unphysical, and we also crack it fully in a variety of ways while we demonstrate that the KLJN scheme stays unbroken as a consequence of the Second Law of Thermodynamics and of noise.

**1.2.4 Bennett–Riedel's wave-transient-based attack before the steady state is reached**

BR write [4]: *Thus, while the steady state mean square noise voltage in the original KLJN protocol does not allow Eve to distinguish between the LH and HL settings of Alice's and Bob's resistors she can distinguish them using (a) transient waves created by the switching action before the steady state is established.*

*For example Bob's resistor affects the phase and amplitude correlations between a right-traveling wave at time t and its left-traveling echo at time t + Δ, where Δ is the transit time from Eve to Bob and back, with the echo vanishing only if the resistor is perfectly impedance matched to his end of the line.*

Here it should be noted that BR have not put forward any concrete protocol with a quantitative and testable evaluation scheme. This is unfortunate because, by establishing such a protocol, one can see that, in the no-wave limit, such transients would represent minor information for Eve about Alice's and Bob's status. Even if propagating signal components (not waves) could be measured, the limited information about the noise within a small fraction of its correlation time (and the unknown additive noise and reflection at the other end of the wire) would make the information available to Eve very small. Moreover, even this minuscule information would converge towards zero upon a decrease of the noise bandwidth and/or a reduction of the wire length. The statistical distance between the key guessed by Eve and that of the perfectly secure key (of the same length) will vanish in a fashion similar to the one described by Eqs. (14) and (15). Sections 2.6 and 2.7 below discuss an efficient transient protocol and quantitative analysis.

**1.2.5 Bennett–Riedel's passive time-correlation attack in the no-wave limit**

BR write [4]: *Thus, while the steady state mean square noise voltage in the original KLJN protocol does not allow Eve to distinguish between the LH and HL settings of Alice's and Bob's resistors, she can distinguish them using (b) time correlations in the steady-state distribution of traveling waves resulting from the fluctuations that give rise to Johnson-Nyquist noise. For example Bob's resistor affects the phase and amplitude correlations between a right-traveling wave at time t and its left-traveling echo at time t + Δ, where Δ is the transit time from Eve to Bob and back, with the echo vanishing only if the resistor is perfectly impedance matched to his end of the line.*

We will analyze this problem in Secs. 2.6 and 2.7 and give a security proof showing that the statistical distance between the key guessed by Eve and the perfectly secure key will vanish in an exponential fashion versus the length of the key.

**1.2.6 Current extraction/injection based active (invasive) attack**

BR write [4]: *...she (Eve) could still learn the key by an active steady-state attack in which she would place a very high-resistance shunt between her node and ground, and monitor the direction of current flow into it. Of course Alice and Bob could try to detect this weak leakage current also, and abort the protocol if they found it. The result would be an unstable arms race, won by whichever side had the more sensitive ammeter, not the sort of robustness reasonably expected of a practical cryptosystem.*

We observe that this attack is valid only against the BR system because, in the KLJN scheme, the direction of the current flowing into the shunt resistor does not provide any information since its origin is a Gaussian noise process with zero mean and exhibiting perfect symmetry around zero. What BR might want to say for the KLJN scheme is that, by using the shunt resistor in the middle, the change of the RMS current in the wire will be greater in the direction of the lower resistance than in the directions of the higher resistance.

A very small difference in current, such as the one referred to above, results in an extremely poor statistics for Eve, and therefore one of the present authors (LK) has proposed a more efficient attack of the mentioned type in the original paper describing the KLJN scheme [26]: this attack entails a separate noise current generator instead of a shunt resistance as well as an evaluation of the cross-correlations between the injected current and the channel currents at the two sides of the injection. These cross-correlations determine which end of the wire has the low and which one has the high resistance. An attack of this type was disregarded as being inefficient already in the foundation paper for KLJN [26], because Eve would need a very long time to create sufficient statistics to reach a reasonable decision, whereas she only has the short duration of the BEP before the process ends. In Sec. 2.8, we analyze this attack mathematically and give a security proof against it.

## Discussion and Results

The flow of analysis and argumentation in this chapter is as follows: First, in Sec. 2.1, we survey well-know facts about the physics related to the no-wave (quasi-static) limit of electrodynamics as well as facts about information transfer in that limit. Then, in Sec. 2.2, we refute BR's claim that there is no information transfer in the quasi-static (no-wave) limit. In Sec. 2.3 we then analyze BR's equations (Eqs. 16 and 17 above) indicating zero security and show that they are invalid for the KLJN scheme not only in the no-wave limit but also in the wave limit, whereas they are indeed valid for BR's thermodynamics-free system. In Sec. 2.4 we demonstrate that BR's thermodynamic-free key exchanger is unphysical because transients will oscillate for infinite time in the wire. Subsequently, in Sec. 2.5, we analyze the real, physical BR system and present ten different ways to fully crack it. We also show there that none of these ways of cracking work against the KLJN scheme, which proves that thermodynamics is essential for the security of KLJN. In Sec. 2.6 we argue that BR are incorrect when they write that the wave-transient attack would crack the KLJN system, and we also find that the statistical distance between the key guessed by Eve and the ideal key exponentially converges zero versus the length of the key. In Secs. 2.7 and 2.8, we demonstrate why BR's passive-correlation attack does not work in the KLJN system and why BR's current-extraction attack fails to change the exponential convergence of the statistical distance to zero. Finally, Sec. 2.9 contains some general remarks about protection against hacking.

## 2.1 Physics facts: Information, propagation, and wave couplers in the quasi-static limit

In Secs. 2.1.1 to 2.1.4 we clarify what is meant by a wave in physics: what are the conditions for the existence of a wave, and what is quasi-static electrodynamics [50] represented by circuit symbols? We also discuss whether electronic circuits are able to transfer signals and information in the quasi-static (no-wave) limit, and we treat the nature of delayed signal propagation in the no-wave limit as well as the inefficiency to separate propagation directions with directional couplers [51]. Here we emphasize it again: the issue is fundamental limitation due to the laws of classical physics.

### 2.1.1 The mathematical definition of a wave in physics

In physics, a wave is defined as a propagating amplitude disturbance $U(x,t)$ that is the solution of the wave equation

$$c^2 \frac{\partial^2 U(x,t)}{\partial x^2} = \frac{\partial^2 U(x,t)}{\partial t^2} \quad , \tag{18}$$

where $c$ is the phase velocity, *i.e.*, the propagation velocity when no dispersion is present. The dynamics of waves is governed by the oscillation of energy between two types, such as the electrical and magnetic field energies. If only one of these types of energy takes part in the propagation—or if the propagation is not based on the bouncing of energy between these two fields—then the propagating field disturbance is not a wave but merely a near-field oscillation with retardation effects.

We now consider a wire with finite size $L$. The wave equation in Eq. (18) has solutions only for frequencies

$$f \geq f_m = \frac{c}{2L} \quad . \tag{19}$$

In other words, propagating field disturbances with frequency components below the minimum wave frequency $f_m$ do not satisfy the wave equation, and hence they are not waves. We concur with BR that propagation and corresponding time delays (*i.e.*, retardation) are essential notions, but the propagating entities are not waves but field relaxations, and the consequences of this will be outlined below. Thus BR's statements about propagating "orthogonal" wave components that can be separated in the two directions is simply unphysical and leads to incorrect equations and conclusions. Furthermore, when the KLJN scheme operates in the "no-wave limit", this means that the condition

$$f \ll f_m = \frac{c}{2L} \tag{20}$$

applies [2,26], and BR are correct in using the term "quasi-static" to describe this situation. However, in the limit of quasi-static electrodynamics [50] it is incorrect to classify the propagating disturbances as waves; these disturbances are neither the solution of the wave equation nor do their electrical and magnetic fields have wave energy bouncing back and forth between them during propagation.

**2.1.2 The quasi-static limit of electrodynamics, and electrical circuitry symbols with lumped elements**

Quasi-static electrodynamics [50] and Eq. (20) constitute the bases for the operation and associated circuit drawings of any electrical circuit with lumped elements. The physical implication is that—along a line in a circuit drawing and the corresponding wire in the realized circuit, and at a given moment—the instantaneous current and the voltage amplitudes are virtually homogeneous, and retardation effects (including waves) can be neglected. In the absence of these implications, everyday electrical engineering design of circuits with lumped elements would be invalid and impossible.

**2.1.3 Signal propagation in the no-wave (quasi-static) limit**

After the comments above it is obvious that BR's assertion, that without waves in the wire there is no information transfer, is not only unphysical but also in blatant contradiction with everyday experience. No landline phones, no computers or other electrical circuits with lumped elements would be able to function and process information if BR's claim were true! In conclusion, the quasi-static (no-wave) limit [50] is a physically valid working condition for the KLJN system, and it is not unphysical as BR claim.

**2.1.4 Further implications of the quasi-static (no-wave) limit: Directional couplers,** *etc*

We now consider *wave-based* directional couplers for extracting and separating signal components in two directions. These couplers simply do not work in the quasi-static limit, and even in the wave limit the cancellation of the irrelevant signal component is strongly frequency dependent because it is determined by the successful destructive interference of wave components in the coupler [51]. Couplers with good directivity are of the size $\lambda_0 / 4$, where $\lambda_0 = c / f_0$ and $f_0$ is the frequency for optimal operation. For longer wavelengths (*i.e.*, smaller frequencies), the system is subject to Rayleigh scattering and, accordingly, the separation of intensities decays with a power function scaling according to $f^4$.

There are also *non-wave-based* directional couplers, which are able to separate signals coming from two directions in the wire. These couplers work with lumped elements, such as transformers or active devices, and can be efficient in a wide frequency range. Their working principle is to cancel the signal of the irrelevant direction by subtracting from the channel voltage another voltage that is induced by the channel current. However, all of these couplers fail with the KLJN key exchanger because, for a proper operation to reveal Alice's voltage spectrum, the designer must know the exact value of her resistor. If instead Bob's resistor value is used, then the resulting signal voltage will be different and signal's spectrum will match Bob's noise spectrum instead. This fact is again a consequence of the Second Law of Thermodynamics, which guarantees that the cross-correlation of the channel voltage and channel current is zero, which leads to statistically independent channel voltage and current as a result of their Gaussian nature. Similarly, measuring the channel voltage $U_c(t)$ and current $I_c(t)$ and creating $U_L^*(t) = U_c(t) \pm I_c(t) R_L$ and $U_H^*(t) = U_c(t) \pm I_c(t) R_H$ would not offer information as a consequence of the independence and the Gaussianity of $U_c(t)$ and $I_c(t)$. According to basic noise calculus, the spectrum of $U_L^*(t)$ and $U_H^*(t)$ would be $4kT_{eff}R_L$ and $4kT_{eff}R_H$, respectively, independently of the sign for the second terms in these sums. In conclusion, non-wave-based directional couplers do not provide useful information for Eve.

## 2.2 Refutation of Bennett–Riedel's claim about no information transfer in the no-wave limit

As already shown in Sec. 2.1.3, there is indeed information transfer in the no-wave limit, and this fact is supported by common experience; *cf*. Eqs. (18) to (20). Therefore, the quasi-static limit is physical in an information processing system.

## 2.3 Invalidity of Bennett–Riedel's equations, and the correct equations

Below, we show that BR's equations are invalid for the KLJN scheme, in the wave limit as well as in the no-wave (quasi-static) limit.

### 2.3.1 The wave limit and the Pao-Lo Liu key exchange system

It is important to note that the default operation of BR's system (*cf*., Fig. 4) is within the wave limit, which is a consequence of the abrupt switching of the voltage (*cf*., Sec. 2.4) and the generated high-frequency products. Moreover, in the BR system, no noise unknown by Eve is fed by Alice and Bob into the system. Thus Eqs. (16) and (17) are indeed valid for the BR system (but not for KLJN). As a consequence, the BR system does not offer any security for Alice and Bob, as further discussed in Sec. 2.5.

The wave limit represents an illegal operational condition for the KLJN scheme, and therefore it is unimportant. However there is a software-based protocol working in the wave limit, known as the Pao-Lo Liu key exchange system [52–54], which was inspired by KLJN but does not utilize the Second Law of Thermodynamics. In the Liu protocol, random number samples of infinitesimally low noises (in the ideal situation) at Alice's and Bob's sites are sent and reflected with random sign of the reflection coefficient. Alice's reflection coefficient, and the noise intensity added by her, is chosen so that, in the steady-state mode of ideal conditions, BR's proposed correlation attack [4] between the incoming and outgoing waves does not yield any information for Eve. The relevant relation for the Liu protocol, in the ideal situation, is

$$H(X|Z_A) = H(X|Z) = H(X) > 0 \tag{21}$$

instead of the zero-security situation, $H(X|Z_A) = H(X|Z) = 0$, implied by BR's considerations and Eq. (17) [4]. Furthermore and surprisingly, Liu's system seems to satisfy

$$I(X;Y|Z) = H(X|Z) - H(X|Z,Y) > 0 \tag{22}$$

in steady-state and at the ideal limit. Liu's system has other weaknesses, though, stemming from the wave limit, *viz.*, the distinct possibility to observe $Z_A$ and $Z_B$; these flaws lead to problems with transients [53] and vulnerability to non-ideal filters [54]. Neither is Liu's system protected by the Second Law of Thermodynamics or other laws of physics.

Finally, returning to the KLJN scheme but lingering in the wave limit, we have the following comments: If only the waves coming from Alice's direction and denoted $Z_A$ are known, this particular situation provides less information about Alice's total description than the situation when the waves $Z_B$ coming from the direction of Bob are known as well. This is so because $Z_A$ alone offers limited information about the reflection coefficient, and the resistance determining it, at Alice's side. On the other hand, in accordance with BR's passive correlation attack discussed in Sec. 1.2.5 (and also in Sec. 2.7 for the no-wave limit, where it does not work), the cross-correlation of $Z_A$ and $Z_B$ (requiring the wave limit) provides more information about the reflection coefficient at Alice than $Z_A$ does, and thus $H(X|Z_A) > H(X|Z)$. We note, in passing, that BR's attack and its justification contradict their own equation, given in Eq. (16), which claims that adding $Z_B$ to the knowledge of $Z_A$ does not help Eve. The duration of the BEP is limited in the KLJN protocol, and thus the relation $H(X|Z_A) > H(X|Z) > 0$ applies in Eq. (21).

It should be observed that Liu's system [52–54], described above, is slightly different from the KLJN system in the wave limit (which is illegal for KLJN) because, in the Liu protocol, the added and reflected noises are combined at the two ends in such a way that, in the ideal case, the cross-correlation does not yield any information from Eve. Thus Liu's system implies that, in general, the correct relation for the wave limit is $H(X|Z_A) \geq H(X|Z) > 0$.

### 2.3.2 Bennett–Riedel's equations for the KLJN scheme in the no-wave (quasi-static) limit

BR's relations in Eqs. (16) and (17) do not exist for the KLJN system in the quasi-static limit because $Z_A$ and $Z_B$ are not observable separately [51]. Directional couplers that are able to separate such waves would produce outputs corresponding to

$$Z'_A = Z_A + (1-\kappa)Z_B \tag{23}$$

and

$$Z'_B = Z_B + (1-\kappa)Z_A \tag{24}$$

with $\kappa \propto 1/f^2$. The largest separation would be at the high cut-off frequency $B_{kljn}$ of the noise bandwidth. As already pointed out in Sec. 2.1.4, this will lead to unconditional $\varepsilon$-security $\left(\varepsilon \propto B_{kljn}^4\right)$, i.e., results of the same nature as in Eqs. (14) and (15). The resources invested by Alice and Bob are the duration $\tau$ of the BEP and the length of the key $\left(\varepsilon \propto 2^{-N}\tau^{-4}\right)$.

Finally, we set up the correct relations replacing Eqs. (16) and (17): The conditional information terms for the KLJN scheme satisfy

$$H(X) > H(X|U_c,I_c) > H(X|U_c,I_c,Z_A^*) \gg H(X|U_c,I_c,Z_A^*,Y) > 0 \quad , \tag{25}$$

where $U_c(x,t)$, $I_c(x,t)$ are current and voltage amplitudes along the wire in the steady state, where the dependence on *x* is miniscule and approaches zero for $B_{kljn} \to 0$, and $Z_A^*(x,t)$ is the initial transient disturbance (not wave) running from Alice toward Bob until Bob's end is reached and Bob's unknown noise is mixed into it. The last conditional information term expresses the fact that Bob, by knowing his own total description, is able to make an almost perfect guess of Alice's description *X* [31]. However this term is still larger than zero, because errors remain even in this case [31,34], implying that a small uncertainty is left. Correspondingly, instead of Eq. (17), the correct relations for the conditional mutual information satisfy

$$I(X;Y|U_c,I_c,Z_A^*) = H(X|U_c,I_c,Z_A^*) - H(X|U_c,I_c,Z_A^*,Y) \gg 0 \quad . \tag{26}$$

### 2.4 Proof that Bennett–Riedel's key exchanger is unphysical

It is easy to see that BR's key exchanger is unphysical in its present form (*cf.*, Fig. 4). To this end, let us consider how long Eve has to "graciously wait" for the termination of the switching transients before she can measure. This time, in fact, is infinite because the transient will bounce back from the two endpoints of the wire, with the same sign from the open end and with altered sign from the endpoint terminated by the battery.

The observations above serve as a clear proof that, in the absence thermodynamics and the loss/energy dissipation it implies, even BR's key exchanger cannot function, and this holds true also if we permit violations of the basic rules of security—*viz.*, that Eve is allowed to measure whenever she can and wants—and instead force Eve to wait until the transients decay, which takes infinitely long time.

In conclusion, BR's scheme is unphysical, and one must realize that there are losses in "real" physical systems and that the related energy dissipation is controlled by thermodynamics.

### 2.5 Ten ways to crack Bennett–Riedel's key exchanger by passive attacks

Below, we show ten ways to crack BR's thermodynamics-free system with 100% success rate, and we furthermore point out that the same cracking methods do not work with the KLJN scheme, which is a consequence of thermodynamics and noise.

### 2.5.1 Six universal energy/current-flow-analysis attacks

To circumvent the problem of waves, Alice and Bob use proper voltage envelopes to prevent wave-modes (high-frequency components belonging to the wave limit). In practice, the wave modes should be kept negligible. Another alternative is that Alice and Bob use filters. One should note that convergence requires some loss, which is unavoidable for any real physical system.

Six universal energy/current-flow-analysis attacks are based on the fact that any wire has a geometrical capacitance, and to charge the wire one needs a current flow, energy flow and power flow. Measurement of voltage and current, and determination their product, gives the power flow and its direction as shown in Fig. 5. This power flow is the quasi-static analogue of the Poynting vector in electromagnetics.

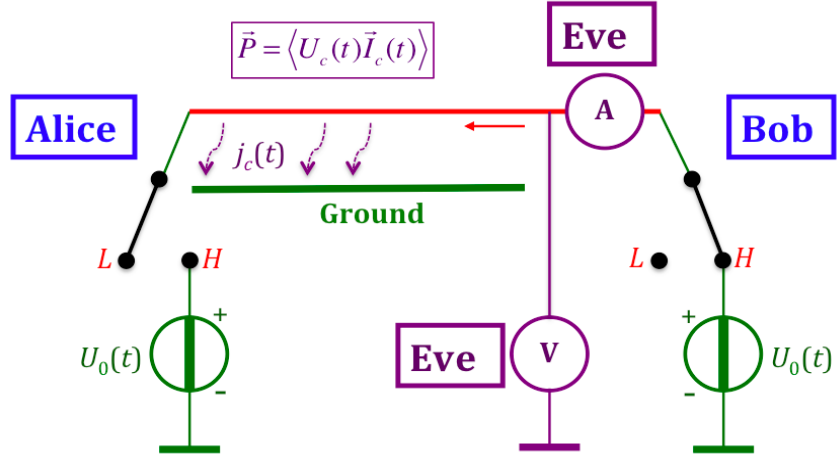

**Figure 5. Universal energy-flow-analysis attack against BR's scheme in the no-wave limit.** The no-wave limit is provided by the specific time-function of the voltage $U_0(t)$. The capacitive current density $j_c(t)$ toward the ground is spatially homogeneous along the wire, which leads to a maximum channel current amplitude $\vec{I}_c(t)$ power flow vector and energy flow vector at the closed end, and zero at the open end. The direction of these vectors during the charge-up period is pointing toward the open end.

The power flow vector is given by

$$\vec{P}(t) = U_c \vec{I}_c(t) \ , \tag{27}$$

and the energy flow vector is its integral over the BEP according to

$$\vec{E} = \int_0^\tau \vec{P}(t) dt \quad . \tag{28}$$

Eve's situation is fully characterized by the direction of the current vector $\vec{I}_c(t)$, the mean power flow vector $\langle \vec{P}(t) \rangle$, and the energy flow vector $\vec{E}$. The magnitudes of the $\langle \vec{P}(x,t) \rangle$, $\vec{E}(x)$ and $\langle \vec{I}(x,t) \rangle$ vectors with regard to location also fully inform Eve and compromise security. The further away from the connected voltage source these location-dependent quantities are evaluated, the less are their values, and they are zero at the open end. The directions of these vectors during the charge-up period are toward the open end.

In conclusion, the direction and the location-dependence of the three measurable quantities offer six ways to fully crack the key in the BR system.

### 2.5.2 Three transient-damping resistor attacks

To make the system physical and stop the transient after one return, Alice and Bob may use damping resistors to match the wave resistance of the wire, as shown in Fig. 6. These resistors will cause a continuous noise current flowing into the geometrical capacitance of the wire. There are then three more ways to utilize thermodynamics to crack this system during the steady state.

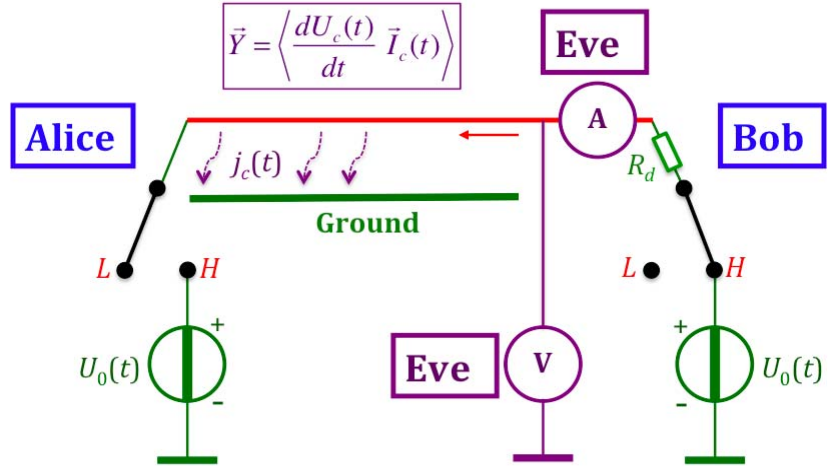

**Figure 6. Transient-damping-resistor version of BR's scheme, and capacitive noise current attack.** The direction and the location-dependent value of the cross-correlation vector $\vec{Y}$ for the time-derivative of the channel voltage and the current vector provide two ways to crack the key, while the location-dependence of the RMS channel current offers a third way.

The noise current is correlated with the time derivative of the channel voltage and can be written

$$\vec{Y} = \left\langle \frac{dU_c(t)}{dt} \vec{I}_c(t) \right\rangle .  \tag{29}$$

Both the sign of the cross-correlation vector $\vec{Y}$ and its value with regard to location fully inform Eve about the situation; their absolute values are zero at the free end of the wire and maximum at the closed end.

A third way to crack the key is given by the location-dependence of the RMS channel current, which is zero at the open end and maximum at the closed end.

### 2.5.3 Wire-resistance Johnson-noise attack

Any wire will have non-zero resistance, and thus it produces Johnson noise. Eve can simply measure the voltage noise between the wire and the ground at the two ends of the wire, as indicated in Fig. 7.

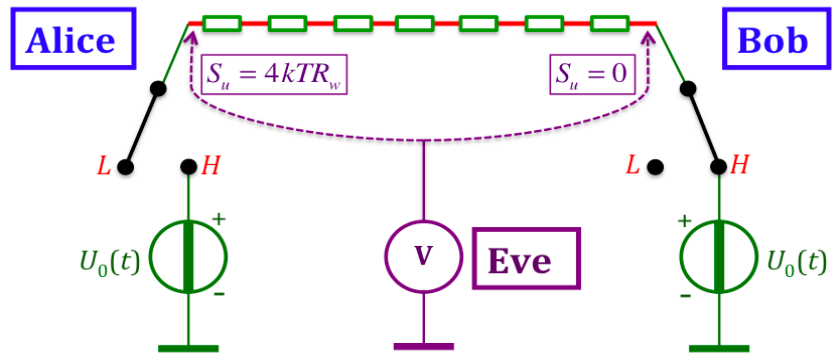

**Figure 7. BR's scheme with finite wire resistance and Johnson noise attack.** With a wire resistance of $R_w$ and in the steady-state mode, Eve will measure a zero-power density spectrum $S_u$ at the closed end of the wire and $S_u = 4kTR_w$ at the open end.

The free end of the wire will have a voltage noise spectrum given by

$$S_u = 4kTR_w ,  \qquad (30)$$

while the connected end shows zero noise. Consequently Eve can fully crack the system. One may note that this attack can be avoided if the connected end of the wire has a large additive noise to conceal the noise given by Eq. (30), but then the former attacks utilizing the current, power flow and energy flow vectors will still crack the system even in the steady state.

### 2.5.4 The above attacks are inefficient against the KLJN system as a result of thermodynamics

It is easy to understand how thermodynamics and noise, fed by the two communicating parties, protect the KLJN scheme against the above attacks. The resistors used by Alice and Bob make the system thermodynamic and produce Johnson noise. The noise voltages are much larger than the parasitic Johnson noise of the wire, because the wire resistance must be small (maximum 1 to 2% of $R_L + R_H$). Similarly the noise bandwidth is chosen so that the capacitive currents are negligible compared to the channel current.

The implication of the considerations above is that, when the above described attacks are used against the KLJN scheme, Eve's measurement will be a small DC signal buried in a large noise. This leads to relations similar to those shown for the wire resistance voltage drop in Eqs. (12) to (15), and the information theoretic security will be almost perfect. In the case of analogous attacks against the BR case, on the other hand, there is no other type of noise to bury Eve's signal. The rectified noise voltages and currents, and the cross-correlation results, are all unipolar noises for which either the polarity of this quantity provides the result or, when its size matters, the size compared to zero. For example, the Johnson noise of the wire should be evaluated only at its two ends and should be zero at one end and non-zero at the other. Neither statistics nor averaging is needed to crack the BR system with these attacks; the result is virtually instantaneous and within the correlation time of the noise.

## 2.6 On transient attacks against the KLJN scheme

This attack is different from other attacks in the literature and in this paper in the sense that, in the no-wave (quasi-static) limit where KLJN operates, no concrete realization has ever been proposed with a measurement and evaluation protocol. Therefore, at the moment, this attack is only hypothetical but is brought up here for the sake of completeness and debate.

Researchers working with the KLJN scheme have realized from the very beginning that transients pose vulnerabilities, and various schemes have been proposed to reduce the potential information leak; they include ramping up/down of the noise, starting from zero noise amplitude (and velocity), and adiabatic random walking of Alice's and Bob's resistance [31]. The efficiency of any attack is strongly limited; firstly, this is due to the quasi-static condition that is manifested by the noise voltage envelope and filters and, secondly, it is due to the unknown resistances and noises at the two ends of the wire. As soon as the front of the propagation (not wave) of the band-limited noise reaches the other end, new information about the particular noise is strongly reduced. Even in the case of a no-transient protocol, Eve has effectively only a very small sample of a noise whose duration is much shorter than its correlation time.

First we describe the so far best-known transient protocol, which is based on random-walk resistances [31]. Alice and Bob arbitrarily choose $R_L$ or $R_H$ as their $R_A$ and $R_B$, and they use continuously variable resistors—such as potentiometers, MOSFETs, *etc*—to execute the key exchange. If noise generators are employed to enhance the noise temperature, then their band-limited white noise spectra also need to be variable in a synchronized fashion so that the noise temperature stays constant at the publicly agreed value $T_{eff}$. Furthermore, suppose that the noise bandwidth in the KLJN scheme is secured by line filters at Alice's and Bob's ends. At the beginning of the KLJN clock period, both Alice and Bob start with

$$R_A(0) = R_B(0) = \frac{R_L + R_H}{2} \quad , \tag{31}$$

and they stay at this value until the noises equilibrate in the wire. Thus no informative transients can be observed just after connecting the resistors to the line, because the bit values have not yet been realized. Then Alice and Bob execute independent, adiabatically slow continuum-time random walks with their resistor values (in a fashion synchronized with the spectral parameter of their noise generators). The random walks are performed so slowly that—from a thermodynamic point of view—the system is changing in the adiabatic limit; thus there is almost thermal equilibrium in the wire during the whole random-walk process.

There is a publicly pre-agreed time $t_r$ to execute these independent random walks. If Alice and Bob reach their randomly preselected values $R_A$ and $R_B$ within this time period, then they stop the random walk and stay at this value. After the time period $t_r$ they restart the measurements in the regular fashion. This procedure virtually removes the transient effects and the information leak they may cause.

If, by the end of the time $t_r$, the random walk of Alice or Bob (or both) does not reach the randomly preselected resistance value, then he/she (or both) submit a cancellation signal via an authenticated channel, and the bit exchange process is immediately aborted; then a new independent KLJN-clock-period starts in the way described above.

Concerning security, the production of spurious frequency products is proportional to the RMS speed $v_{rms}$ of the random walk and, if a concrete attack is implemented, it is reasonable to assume that it satisfies

$$q = \vartheta_{tr} v_{rms} , \qquad (32)$$

where $\vartheta_{tr}$ is a constant relevant for the transient attack against this scheme (*cf*. Eqs. 12 and 13). The above assumption leads to unconditional $\varepsilon$-security $(\varepsilon \propto v_{rms})$ with results of the same nature as those given in Eqs. (12) and (15) and with statistical distance

$$\Delta = (0.5 + q)^N - 0.5^N \cong 2Nq0.5^N = 2N\vartheta_{tr} v_{rms} 0.5^N . \qquad (33)$$

If $q$ is not small enough, it can further be reduced by the privacy amplification [33] because of the high fidelity of the KLJN scheme.

Here the resource used to approach the perfect security is the duration $\tau$ of the BEP, because it is inversely proportional to $v_{rms}$ when the random walk time is dominating. In other words, at fixed key length the "price" of increasing the security is a reduction of the speed of key exchange, and $\varepsilon \propto 0.5^N \tau^{-1} N$ can again be arbitrarily small.

## 2.7 Why Bennett–Riedel's passive correlation attack does not work against KLJN

Directional couplers have limited bandwidth, work in the wave limit and—given that their directivity is good—have a size $\lambda_0 / 4$ (*cf*. Sec. 2.1.4). For much longer wavelengths—*i.e.*, smaller frequencies, as in the KLJN scheme—the system displays Rayleigh scattering and accordingly (*cf*. Eqs. 23 and 24) the passive correlation attack results in a correlation coefficient with power function scaling according to $f^4$. These conditions lead to unconditional $\varepsilon$-security $\left(\varepsilon \propto f^4\right)$ with results that again are of the same nature as those in Eqs. (12) and (15) and with statistical distance

$$\Delta = (0.5+q)^N - 0.5^N \cong 2Nq0.5^N = 2N\vartheta_{cr} B_{kljn}^4 0.5^N ,  \qquad (34)$$

where $\vartheta_{cr}$ is a constant defined as in Eq. (13). This value can further be enhanced by privacy amplification [33], if needed.

Here the resource used to approach perfect security is the duration $\tau$, $\left(\tau \propto 1/B_{kljn}\right)$ of the BEP, because it is inversely proportional to the highest frequency in the noise-bandwidth. In other words, at fixed key length the cost of increasing the security is a reduction of the speed of key exchange, and $\varepsilon \propto 0.5^N \tau^{-4} N$ can again be arbitrarily small.

**2.8 Why the current extraction/injection active attack does not work against KLJN**

BR [4] propose an active (invasive) attack wherein Eve connects a grounded resistor to the line in order to extract some current and also monitors the current direction in the wire. BR's verbatim statement was reproduced in Sec. 1.2.6 above.

We fully agree with the above assessment when it refers to the BR system. However this attack is inefficient against the KLJN scheme, and this fact was pointed out already in the original paper describing the KLJN scheme [26]. In fact, this latter paper proposes a technically more efficient attack of the same nature: that Eve injects a stochastic current at the middle and monitors the cross-correlation of this current with the channel currents in the two directions; the correlation coefficient will be greater in the direction of the smaller resistance. This attack was later pointed out also by Reiner Plaga and Horace Yuen in private communications. Alice and Bob monitor the channel currents at the two ends and compare their instantaneous amplitudes via an authenticated public channel. If the currents differ, the bit exchange event is terminated and that bit is discarded.

In this section we show a mathematical proof that the *uncertainly principle between measurement duration and statistical error*s makes it impossible for Eve to crack the key and the unconditional security remains even against this type of attack. The usual argument to justify the attacks referred to above is that Eve may use miniscule current amplitudes, which are below the detection limit of the comparisons by Alice and Bob. This argument does not work, however, because Alice and Bob can design their current resolution so that Eve, by implementing this attack, cannot extract enough information. Mathematically, the channel current at Alice's side of Eve is

$$I_{cA}(t) = I_c(t) - \gamma I_E(t) , \qquad (35)$$

and at Bob's side of Eve it is

$$I_{cB}(t) = I_c(t) + (1-\gamma)I_E(t) , \qquad (36)$$

where $I_E(t)$ is Eve's injected current and $(1-\gamma)/\gamma = R_A/R_B$. The cross-correlations with Eve's current during the BEP are

$$\rho_A = \left\langle [I_c(t) - \gamma I_E(t)] I_E(t) \right\rangle_\tau = \left\langle I_c(t) I_E(t) \right\rangle_\tau - \gamma \left\langle I_E^2(t) \right\rangle_\tau = U_{cE\tau}(t) - \gamma \left\langle I_E^2(t) \right\rangle - \gamma U_{EE\tau}(t) , \quad (37)$$

$$\rho_B = \left\langle [I_c(t) + (1-\gamma) I_E(t)] I_E(t) \right\rangle_\tau = \left\langle I_c(t) I_E(t) \right\rangle_\tau + (1-\gamma) \left\langle I_E^2(t) \right\rangle_\tau =$$
$$= U_{cE\tau}(t) + (1-\gamma) \left\langle I_E^2(t) \right\rangle + (1-\gamma) U_{EE\tau}(t) , \quad (38)$$

where $\langle \ \rangle_\tau$ stands for finite-time ($\tau$) average, $U$ for noise components, and $\langle \ \rangle$ for the exact average (requiring infinite time). The dominant terms at the right-hand side of Eqs. (37) and (38) are the noise terms of the cross-correlations between Eve's current and the channel current, with mean-square amplitudes scaling as $\tau^{-1}$. The RMS amplitude $I_{E,rms}$ of Eve's current is negligible compared to that of the channel current, and hence

$$I_{E,rms} = \sigma I_{c,rms} , \quad (39)$$

where $\sigma \ll 1$. The last noise terms at the right-hand side of Eqs. (37) and (38) are negligible compared to the first noise terms. The detection problem is again the same as the one encountered at the wire-resistance-attack: a small DC component (the second term) in a large noise (the first term). Thus, as inferred from Eqs. (12) and (13), $q$ will again satisfy

$$q = \vartheta_{ci} \sigma , \quad (40)$$

where $\vartheta_{ci}$ is a constant relevant for this current injection/extraction attack at fixed $\tau$ (note that $\vartheta_{ci}$ is inversely proportional to $\tau$). Again one reaches *unconditional $\varepsilon$-security* $\left(\varepsilon \propto 0.5^N \sigma\right)$, with results of the same nature as those in Eqs. (12) to (15) and with statistical distance

$$\Delta = (0.5 + q)^N - 0.5^N \cong 2Nq0.5^N = 2N\vartheta_{ci}\sigma 0.5^N . \quad (41)$$

As a practical example, let us consider 14 bits accuracy of current/voltage comparison defense by Alice and Bob, which means $\sigma < 10^{-4}$. For a similar value of asymmetry during the experimental demonstration [29], the resulting $q$ was 0.025. Thus a two-step XOR-type privacy amplification [33], described in Section 1.1.4 will achieve $q < 5 \times 10^{-5}$, which allows key lengths $N$ up to the order of $10^4$ bits. . Similarly to the situation in Section 1.1.4, for a 1000-bit-long shared key, it gives $\Delta(E,I)_{1000} = 9.3 \times 10^{-303}$ (*i.e.*, an $\varepsilon$-security with $\varepsilon_{1000} \cong 10^{-302}$); for a 500-bit-long shared key it results in $\Delta(E,I)_{500} = 1.5 \times 10^{-152}$, *i.e.*, an $\varepsilon$-security level with $\varepsilon_{500} \cong 2 \times 10^{-152}$.

## 2.9 Remarks about potential hacking attacks

Mathematical models of physical systems and their building elements are always approximate, and security proofs can only be given for these model systems. Particularly dangerous are the elements that are directly exposed to Eve. Thus a commercial secure key exchanger must be carefully designed with considering all the foreseeable hacking attacks.

For example, a real KLJN scheme must be armed with extra circuitry and protocol steps against Makarov-style blinding attacks [11], circulator-based attacks [32], and other unexplored possibilities such as out-of-frequency-range probing attacks, *etc*.

## Conclusions

We showed that thermodynamics, noise, and the Second Law of Thermodynamics—*i.e.*, the impossibility to construct a perpetual motion machine of the second kind—are essential for the security of the classical physical key exchanger in the KLJN scheme. Furthermore we supplied mathematical security proofs for each attack proposed by Bennett and Riedel [4]. Our results indicate that the security of the KLJN system has not been successfully challenged by them.

We also showed that the Bennett–Riedel scheme is unphysical and we cracked it with 100% success by passive attacks, in ten different ways. It was found that the same cracking methods do not function for the KLJN scheme. Some other claims by BR we subjected to critical analysis as well; for example, we proved that their equations for describing zero security do not apply for the KLJN scheme.

It is important to emphasize that all our analyses have assumed a *technically-unlimited Eve* with infinitely accurate and fast measurements. For non-ideal situations and at active (invasive) attacks, the *uncertainly principle between measurement duration and statistical errors* makes it impossible for Eve to extract the key regardless of the accuracy or speed of her measurements.

## Acknowledgements

LK is grateful to Horace Yuen, Vadim Makarov, Renato Renner and Vincent Poor for helpful discussions, consultations and critical remarks about relevant security measures, and to Robert Mingesz, Henning Dekant, John Norton and Gabor Schmera for related discussions. Furthermore, we are indebted to Charles Bennett and Jess Riedel for their work, cited in Ref. [4], which has given us an opportunity to sharpen our arguments in favor of the Kirchhoff-law–Johnson-noise (KLJN) scheme.

## References

1. Liang Y, Poor HV, Shamai S (2008) Information theoretic security. Foundations Trends Commun. Inform. Theory 5:355-580. DOI: 10.1561/0100000036.
2. Mingesz R, Kish LB, Gingl Z, Granqvist CG, Wen H, Peper F, Eubanks T, Schmera G (2013) Unconditional security by the laws of classical physics. Metrology & Measurement Systems XX:3-16. DOI: 10.2478/mms-

2013-0001. http://www.degruyter.com/view/j/mms.2013.20.issue-1/mms-2013-0001/mms-2013-0001.xml
3. Bennett CH, Brassard G (1984) Proc. Int. Conf. on Computers, Signals, and Signal Processing, Bangalore, India. pp. 175-179.
4. Bennett CH, Riedel CJ (2013) On the security of key distribution based on Johnson-Nyquist noise. http://arxiv.org/abs/1303.7435
5. Yuen HP (2012) On the foundations of quantum key distribution—Reply to Renner and beyond. arXiv:1210.2804.
6. Hirota O (2012) Incompleteness and limit of quantum key distribution theory. arXiv:1208.2106v2.
7. Renner R (2012) Reply to recent scepticism about the foundations of quantum cryptography. arXiv:1209.2423v.1.
8. Yuen HP (2012) Security significance of the trace distance criterion in quantum key distribution. arXiv:1109.2675v3.
9. Yuen HP (2009) Key generation: Foundation and a new quantum approach. IEEE J. Selected Topics in Quantum Electronics 15, 1630.
10. Salih H, Li ZH, Al-Amri M, Zubairy MS (2013) Protocol for direct counterfactual quantum communication. Phys. Rev. Lett. 110:170502.
11. Merali Z (29 August 2009) Hackers blind quantum cryptographers. Nature News, DOI:10.1038/news.2010.436.
12. Gerhardt I, Liu Q, Lamas-Linares A, Skaar J, Kurtsiefer C, Makarov V (2011) Full-field implementation of a perfect eavesdropper on a quantum cryptography system. Nature Commun. 2; article number 349. DOI: 10.1038/ncomms1348.
13. Lydersen L, Wiechers C, Wittmann C, Elser D, Skaar J, Makarov V (2010) Hacking commercial quantum cryptography systems by tailored bright illumination. Nature Photonics 4:686-689. DOI: 10.1038/NPHOTON.2010.214.
14. Gerhardt I, Liu Q, Lamas-Linares A, Skaar J, Scarani V, Makarov V, Kurtsiefer C (2011) Experimentally faking the violation of Bell's inequalities. Phys. Rev. Lett. 107:170404. DOI: 10.1103/PhysRevLett.107.170404.
15. Makarov V, Skaar J (2008) Faked states attack using detector efficiency mismatch on SARG04, phase-time, DPSK, and Ekert protocols. Quantum Inform. Comp. 8:622-635.
16. Wiechers C, Lydersen L, Wittmann C, Elser D, Skaar J, Marquardt C, Makarov V, Leuchs G (2011) After-gate attack on a quantum cryptosystem. New J. Phys. 13:013043. DOI: 10.1088/1367-2630/13/1/013043.
17. Lydersen L, Wiechers C, Wittmann C, Elser D, Skaar J, Makarov V (2010) Thermal blinding of gated detectors in quantum cryptography. Opt. Express 18:27938-27954. DOI: 10.1364/OE.18.027938.
18. Jain N, Wittmann C, Lydersen L, Wiechers C, Elser D, Marquardt C, Makarov V, Leuchs G (2011) Device calibration impacts security of quantum key distribution. Phys. Rev. Lett. 107:110501. DOI: 10.1103/PhysRevLett.107.110501.
19. Lydersen L, Skaar J, Makarov V (2011) Tailored bright illumination attack on distributed-phase-reference protocols. J. Mod. Opt. 58:680-685. DOI: 10.1080/09500340.2011.565889.
20. Lydersen L, Akhlaghi MK, Majedi AH, Skaar J, Makarov V (2011) Controlling a superconducting nanowire single-photon detector using tailored bright illumination. New J. Phys. 13:113042. DOI: 10.1088/1367-2630/13/11/113042.
21. Lydersen L, Makarov V, Skaar J (2011) Comment on "Resilience of gated avalanche photodiodes against bright illumination attacks in quantum cryptography". Appl. Phys. Lett. 99:196101. DOI: 10.1063/1.3658806.
22. Sauge S, Lydersen L, Anisimov A, Skaar J, Makarov V (2011) Controlling an actively-quenched single photon detector with bright light. Opt. Express 19:23590-23600.
23. Lydersen L, Jain N, Wittmann C, Maroy O, Skaar J, Marquardt C, Makarov V, Leuchs G (2011) Superlinear threshold detectors in quantum cryptography. Phys. Rev. Lett. 84:032320. DOI: 10.1103/PhysRevA.84.032320.
24. Lydersen L, Wiechers C, Wittmann C, Elser D, Skaar J, Makarov V (2010) Avoiding the blinding attack in QKD: Reply (Comment). Nature Photonics 4:801-801. DOI: 10.1038/nphoton.2010.278.
25. Makarov V (2009) Controlling passively quenched single photon detectors by bright light. New J. Phys. 11:065003. DOI: 10.1088/1367-2630/11/6/065003.
26. Kish LB (2006) Totally secure classical communication utilizing Johnson(-like) noise and Kirchhoff's law. Phys. Lett. A 352:178-182.
27. Cho A (2005) Simple noise may stymie spies without quantum weirdness. Science 309:2148; http://www.ece.tamu.edu/~noise/news_files/science_secure.pdf.

28. Kish LB (2006) Protection against the man-in-the-middle-attack for the Kirchhoff-loop-Johnson(-like)-noise cipher and expansion by voltage-based security. Fluct. Noise Lett. 6:L57-L63. http://arxiv.org/abs/physics/0512177.
29. Mingesz R, Gingl Z, Kish LB (2008) Johnson(-like)-noise-Kirchhoff-loop based secure classical communicator characteristics, for ranges of two to two thousand kilometers, via model-line, Phys. Lett. A 372:978-984.
30. Palmer DJ (2007) Noise encryption keeps spooks out of the loop. New Scientist, issue 2605 p.32; http://www.newscientist.com/article/mg19426055.300-noise-keeps-spooks-out-of-the-loop.html.
31. Kish LB (2013) Enhanced secure key exchange systems based on the Johnson-noise scheme. Metrology & Measurement Systems XX:191–204. open access: http://www.degruyter.com/view/j/mms.2013.20.issue-2/mms-2013-0017/mms-2013-0017.xml?format=INT
32. Kish LB, Horvath T (2009) Notes on recent approaches concerning the Kirchhoff-law-Johnson-noise-based secure key exchange. Phys. Lett. A 373:901-904.
33. Horvath T, Kish LB, Scheuer J (2011) Effective privacy amplification for secure classical communications. Europhys. Lett. 94:28002. http://arxiv.org/abs/1101.4264.
34. Saez Y, Kish LB, Mingesz R, Gingl Z, Granqvist CG (2013) Current and voltage based bit errors and their combined mitigation for the Kirchhoff-law-Johnson-noise secure key exchange. http://arxiv.org/abs/1309.2179. http://vixra.org/abs/1308.0113 .
35. Kish LB, Kwan C (2013) Physical uncloneable function hardware keys utilizing Kirchhoff-law-Johnson-noise secure key exchange and noise-based logic. http://vixra.org/abs/1305.0068; http://arxiv.org/abs/1305.3248
36. Kish LB, Saidi O (2008) Unconditionally secure computers, algorithms and hardware. Fluct. Noise Lett. 8:L95-L98.
37. Gonzalez E, Kish LB, Balog R, Enjeti P (2013) Information theoretically secure, enhanced Johnson noise based key distribution over the smart grid with switched filters. http://vixra.org/abs/1303.0094; http://arxiv.org/abs/1303.3262.
38. Kish LB, Mingesz R (2006) Totally secure classical networks with multipoint telecloning (teleportation) of classical bits through loops with Johnson-like noise. Fluct. Noise Lett. 6:C9-C21.
39. Kish LB, Peper F (2012) Information networks secured by the laws of physics. IEICE Trans. Commun. E95-B:1501-1507.
40. Scheuer J, Yariv A (2006) A classical key-distribution system based on Johnson (like) noise – How secure? Phys. Lett. A 359:737-740.
41. Kish LB, Scheuer J (2010) Noise in the wire: The real impact of wire resistance for the Johnson(-like) noise based secure communicator. Phys. Lett. A 374:2140-2142.
42. Kish LB (2006) Response to Scheuer-Yariv: "A classical key-distribution system based on Johnson (like) noise – How secure?". Phys. Lett. A 359:741-744.
43. Hao F (2006) Kish's key exchange scheme is insecure. IEE Proc. Inform. Soc. 153:141-142.
44. Kish LB (2006) Response to Feng Hao's paper "Kish's key exchange scheme is insecure". Fluct. Noise Lett. 6:C37-C41.
45. Liu PL (2009) A new look at the classical key exchange system based on amplified Johnson noise. Phys. Lett. A 373:901-904.
46. Arora S, Barak B (2009) Computational Complexity. Cambridge University Press, Cambridge.
47. Kish LB (2006). Thermal noise driven computing. Appl. Phys. Lett. 89:144104.
48. Kish LB, Granqvist CG (2013) On the security of the Kirchhoff-law-Johnson-noise (KLJN) communicator. http://arxiv.org/abs/1309.4112 .
49. Antilli D (2005) System and method for the propagation of deterministic and non-deterministic values by means of electrical conductors. Patent publication number EP1952573 (A2). http://www.google.com/patents/EP1952573A2?cl=en
50. Pauli W (2000) Electrodynamics. Dover Publications, New York.
51. Matthaei GL, Young L, Jones EMT (1964) Microwave Filters, Impedance-Matching Networks, and Coupling Structures. McGraw-Hill, New York.
52. Liu PL (2009) A key agreement protocol using band-limited random signals and feedback. IEEE J. Lightwave Technol. 27:5230-5234.
53. Liu PL (2009) Security risk during the transient in a key exchange protocol using random signals and feedback. Phys. Lett. A 373:3207–3211.
54. Kish LL, Zhang B, Kish LB (2010) Cracking the Liu key exchange protocol in its most secure state with Lorentzian spectra. Fluct. Noise Lett. 9:37-45.